# Parity Breaking and Sublattice Dichotomy in Monolayer FeSe Superconductor

Cui Ding,[1,2,#] Zhipeng Xu,[3,4,#] Xiaotong Jiao,[1] Yinqi Hu,[1] Wenxuan Zhao,[1] Lexian Yang,[1,5] Kun Jiang,[3,4,*] Lili Wang,[1,5,†] Jin-Feng Jia,[2,6,7] Jiangping Hu,[3,8,‡] Qi-Kun Xue,[1,2,5,6,§]

[1]State Key Laboratory of Low-Dimensional Quantum Physics, Department of Physics, Tsinghua University, Beijing 100084, China

[2]Quantum Science Center of Guangdong-Hong Kong-Macao Greater Bay Area, Shenzhen 518045, China

[3]Beijing National Laboratory for Condensed Matter Physics and Institute of Physics, Chinese Academy of Sciences, Beijing 100190, China

[4]School of Physical Sciences, University of Chinese Academy of Sciences, Beijing 100190, China

[5]Frontier Science Center for Quantum Information, Beijing 100084, China

[6]Department of Physics, Southern University of Science and Technology, Shenzhen 518055, China

[7]Department of Physics and Astronomy, Shanghai Jiao Tong University, Shanghai 200240, China

[8]New Cornerstone Science Laboratory, Beijing, 100190, China

[#]These authors contributed equally: C. Ding, Z. Xu.

**ABSTRACT**. A unit cell represents the smallest repeating structure in solid-state physics and serves as the fundamental building block of a material. In iron-based superconductors, each unit cell contains two iron atoms, which form two sublattices in the two-dimensional iron layers. Under normal circumstances, these sublattices are expected to have identical physical properties due to space inversion symmetry. However, we discover that this sublattice structure can introduce a novel degree of freedom for probing unconventional pairing mechanisms in iron-based superconductors. We observe distinct dual tunneling spectra within the pairing gap energy corresponding to the two sublattices in the monolayer FeSe with atomically homogeneous (1 × 1) structures on $SrTiO_3(001)$ substrates - a phenomenon we term *sublattice dichotomy*. This dichotomy can be quantitatively explained by a parity-breaking superconducting state, characterized by the coexistence of conventional pairing and interband odd-parity pairing. The interband singlet pairing arises due to the lack of inversion symmetry, which is naturally broken from the interface coupling between FeSe and $TiO_2$ layer.

## I. INTRODUCTION.

Monolayer FeSe grown on $SrTiO_3$(001) is a unique iron-based high-temperature superconductor that boosts Cooper pairs at a record temperature above 65 K [1-4], significantly higher than its bulk counterpart values of 9-14 K and up to 48 K under heavy doping [5-10]. Since it was discovered in 2012 [1], tremendous efforts have been devoted to reproducing, analyzing, and generalizing this unexpected finding [2-4,11-24]. The monolayer FeSe only contains Fermi surfaces around the Brillouin zone corner, whereas common iron-based superconductors have another group of $\Gamma$ hole pockets [2,4,11-14]. How pairing states are established and how the superconducting features deviate from those of other iron-based superconductors is widely debated. However, owing to the diverse interface structures, the physical properties of monolayer FeSe, especially their superconducting nature, remain elusive.

The primitive unit cell of a single iron-based superconducting layer contains two Fe sublattices that bond alternately with upper-layer Se ($Se_+$) and bottom-layer Se ($Se_-$), i.e., α-Fe aligned in $Se_-$-[100] and $Se_+$-[010] and β-Fe otherwise, as illustrated in Fig. 1(a). For monolayer FeSe on $SrTiO_3$(001), the inversion symmetry is broken, for the distances between the $Se_+$-Fe plane and the $Se_-$-Fe plane become asymmetric due to interface coupling [25-27]. Figure 1(b) displays the schematic of the Fermi surface in the 2-Fe Brillouin zone disclosed by angle-resolved photoemission spectroscopy (ARPES) measurement (see Section 2 of the supplementary materials (SM), Fig. S1). As a result of interface charger transfer, there are simply Fermi surfaces around the $M$ points, i.e., the electron pocket with the bottom at - 60 meV, while the $\Gamma$-hole pocket is immersed at ~ 80 meV below the Fermi level ($E_F$) (Fig. S1).

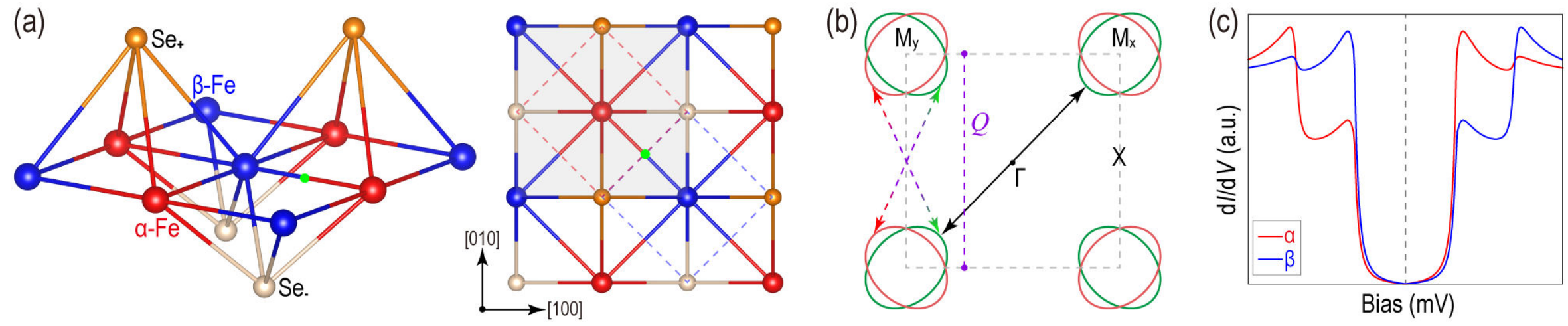

FIG 1. Two-Fe sublattices and mixed interband and intraband pairing in monolayer FeSe. (a) Schematic of the lattice of monolayer FeSe showing two inequivalent sublattices. The Fe atoms aligned in $Se_-$-[100] and $Se_+$-[010] are labeled as α-Fe (red balls), and the other ones as β-Fe (blue balls). The shaded square denotes the primitive unit cell that contains two Fe sublattices, the red and blue dashed square mark 1-Fe unit cells, and the green dot labeled at the midpoint of the adjacent Fe sublattices indicates the position of the inversion operator for the bulk FeSe. (b) Fermi surface based on the 2-Fe Brillouin zone and systematic illustration of the normal pairing between $(k\uparrow, -k\downarrow)$ (black double arrow) and the interband pairing between $(k\uparrow, -k+Q\downarrow)$ (red-green double arrow), where $Q = (\pi, \pi)$ (purple double-dot). (c) Schematic tunneling spectra of α-Fe and β-Fe sublattices showing the sublattice dichotomy.

In theoretical models based on 1-Fe unit cells [28], there are two possible pairing schemes, normal intraband pairing and interband pairing, illustrated by the black and red-green double-arrows, respectively. The normal pairings are zero-total-momentum Cooper pairs between $\boldsymbol{k}$ and $-\boldsymbol{k}$, while the interband pairings are Cooper pairs between $\boldsymbol{k}$ and $-\boldsymbol{k} + \boldsymbol{Q}$, which is allowed as the momentum vector $\boldsymbol{Q} = (\pi, \pi)$ is a reciprocal lattice vector for the genuine 2-Fe unit cell. Notice that the normal pairing and the interband pairing have opposite parity values without mixing. The interband pairing between $\boldsymbol{k}$ and $-\boldsymbol{k} + \boldsymbol{Q}$ was proposed over a decade ago [28], but its consequences remain unexplored. The broken inversion symmetry at the FeSe/$SrTiO_3$ interface allows the coexistence of these two types of pairings. Therefore, the monolayer FeSe offers an opportunity to explore possible novel pairings.

In this work, we carry out a systematic investigation of molecular beam epitaxial (MBE) monolayer FeSe with an atomically homogeneous (1 × 1) surface structure on $SrTiO_3$(001) using atomic-resolution scanning tunneling microscopy/spectroscopy (STM/STS). We find that the tunneling spectra of α-Fe and β-Fe are significantly different, resulting in a sublattice dichotomy in tunneling spectra coherence peaks, as illustrated in Fig. 1(c). More specifically, monolayer FeSe is a two-band system with a dual-gap [1,16,22,29], i.e., one pair of inner-gap coherence peaks at $\pm V_{\mathrm{i}}$ and the other pair of outer-gap coherence peaks at $\pm V_{\mathrm{o}}$. We find that the coherence peak at $-V_{\mathrm{i}}$ for α-Fe is lower than that for β-Fe, whereas the coherence peak at $+V_{\mathrm{i}}$ for α-Fe is higher by a comparable difference. Moreover, the intensity contrast at $\pm V_{\mathrm{o}}$ coherence peaks is opposite to that at $\pm V_{\mathrm{i}}$ for each sublattice. This sublattice dichotomy can be explained by the substantial coexistence of both normal pairing and interband pairing in the superconducting state.

## II. METHODS

### Sample preparation

Our experiments were carried out in a Createc ultrahigh vacuum ($1.0 \times 10^{-10}$ mbar) low-temperature STM system equipped with an MBE chamber. The Nb-doped $SrTiO_3$(001) (0.05 wt.%) substrates were annealed above 1000 °C to obtain dual-$TiO_{2-\delta}$-termination. Monolayer FeSe films were prepared via standard co-evaporation and post-annealing, with a deposition rate of ~0.02 monolayers per minute at a substrate temperature of 470 °C.

### STM experiments

All the STM measurements were performed in constant current mode (tunneling current set point $I_t$ = 500 pA) with the bias voltage applied to the sample ($V_s$), using a polycrystalline PtIr tip. The differential conductance d$I$/d$V$ spectra, which characterize the local density of states around $E_{\mathrm{F}}$, are measured by disabling the feedback circuit, sweeping the sample voltage $V_s$, and then extracting the differential tunneling current d$I$/d$V$ via a standard lock-in technique with a small bias modulation (~1% of the sweeping range) at 937 Hz.

## III. RESULTS

**Fe sublattices identification**

Figures 2(a-d) and 2(e-h) summarize the atomically resolved topography and tunneling spectra $g(r, V) \equiv dI/dV(r, V)$ collected under liquid helium and liquid nitrogen temperatures, respectively, showing the atomic contrast in $Se_-$-Fe-$Se_+$ triple layers. Displayed in Figs. 2(a) and 2(b, c) are the atomic-resolution topographic images taken at sample bias setpoints $V_s$= 80 meV and $V_s$= -50 meV, respectively, presenting the $Se_+$(001) surface. The inset fast Fourier transformed (FFT) images in Figs. 2(a) and 2(b) consistently show sharp $Q_a$ and $Q_b$ Bragg spots corresponding to the (1 × 1) primitive unit cell with in-plane lattice constants of 3.89 Å, while additional $Q_x$ and $Q_y$ Bragg spots of the 1-Fe sublattice emerge in the latter one. Therefore, the monolayer film is FeSe-(1 × 1) that is devoid of any electronic modulation, rather than the ubiquitous (2 × 1) order reported previously [1,16,20]. In the atomic-resolution zoom-in image in Fig. 2(c), the $Se_+$ atom sites, Fe atom sites, and $Se_-$ atom sites are individually discerned, with apparent atomic contrast

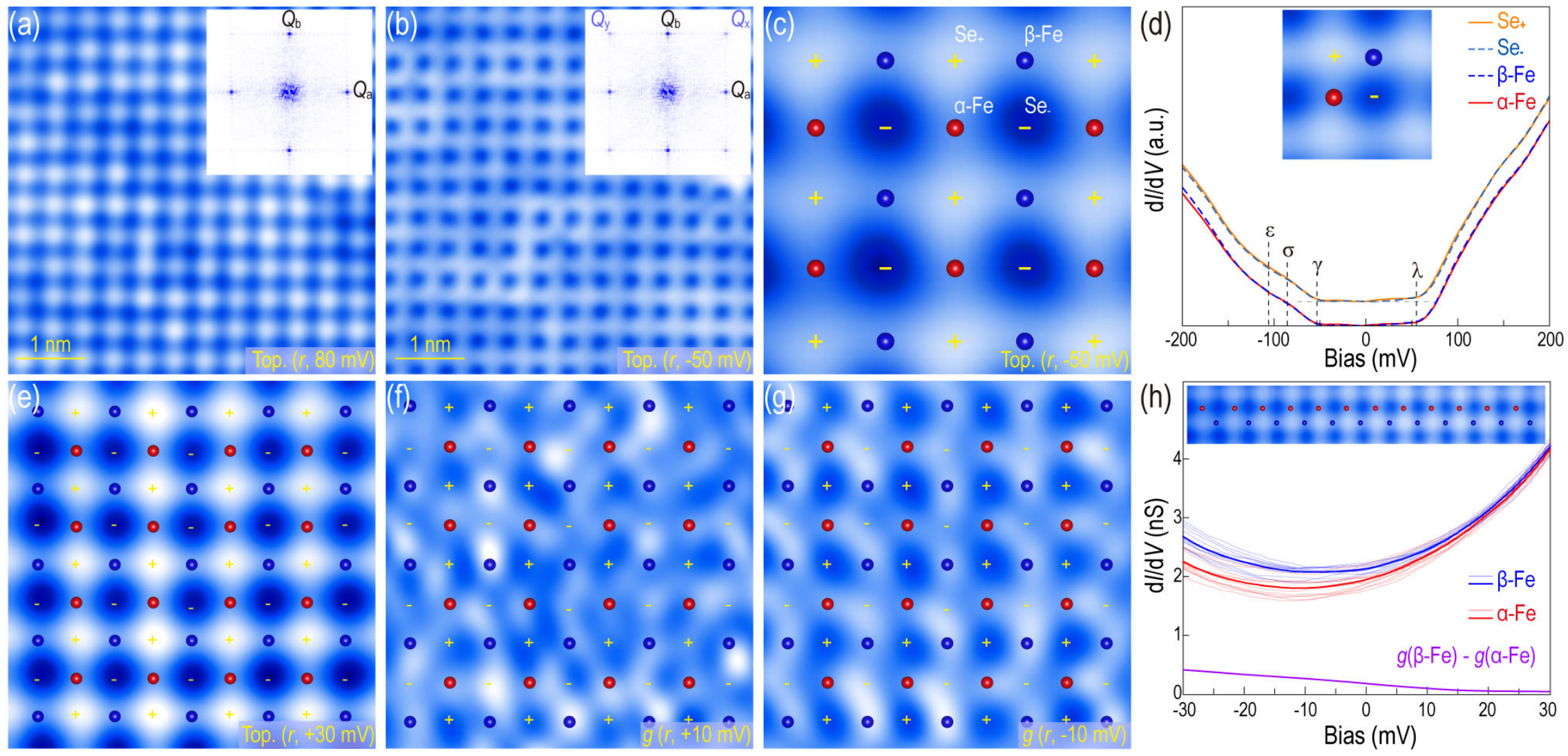

FIG 2. Atomic sites identification of the monolayer FeSe-(1×1) surface under liquid helium temperature (a-d) and liquid nitrogen temperature (e-h). (a) and (b) Bias-dependent atomic-resolution topography, with the corresponding FFT patterns inset, and (c) the zoom-in image. The superimposed red (blue) balls mark α-Fe (β-Fe) sites, and the "+" ("-") mark upper-layer $Se_+$ (bottom-layer $Se_-$) sites. (d) Large-bias tunneling spectra taken at the four Fe/Se atomic sites labeled in the inset. The multiple kinks marked with vertical dashed lines are in alignment with the band structure in Fig. S1. (e) Atomic-resolution topography, and (f, g) simultaneously collected $g(r, \pm 10$ mV) mapping images. (h) Small-bias tunneling spectra taken at the marked Fe sites in the inserted topography, with the respective averages (in bold) and the average difference (in purple). Setpoint: (a), $V_s$ = 80 mV; (b, c), $V_s$ = - 50 mV; (d), $V_s$ = 500 mV; (e, f, g), $V_s$ = 30 mV; (h), $V_s$ = + 50 mV.

matching well with the superimposed lattice model. Plotted in Fig. 2(d) are the typical large-bias tunneling spectra of the four atomic sites in one 2-Fe unit cell collected under bias setpoint $V_s$= 500 meV, showing bias-dependent behavior in alignment with the band structure (Fig. S1). In contrast to the V-shaped tunneling spectra in bulk FeSe [30], the characteristic low tunneling conductance plateau around $E_F$ (± 60 mV) is due to the absence of the $\Gamma$-centered Fermi surface and the low tunneling probability of the $M$-centered electron bands [17]. It is under bias setpoints within the tunneling plateau, i.e., the M-centered orbitals dominate the tunneling, that the ($Q_x$, $Q_y$) vectors for 1-Fe unit cell occur, as exemplified in Fig. 2(b) (see more bias-dependent topography in Fig. S2). Compared with Se sites, the spectra of two Fe sublattices show a slightly flatter tunneling conductance plateau, but with indiscernible contrast between themselves.

The normal-state tunneling spectra collected under liquid nitrogen temperature show minor contrast in the two Fe-sublattices within the energy range of the $M$-centered orbitals. Figures 2(f) and 2(g) present the atomic-resolution tunneling conductance $g(r, V)$ mapping images with $V$ = ± 10 mV, obtained simultaneously with the topography in Fig. 2(e), and Fig. 2(h) summarizes the small-bias tunneling spectra of Fe sites collected under bias setpoint $V_s$= 50 meV. The bottom-layer $Se_-$ sites show significantly higher tunneling conductance than the upper-layer $Se_+$ sites, contrary to the contrast in topography, while the β-Fe sites show a slightly higher conductance than the α-Fe sites. Direct atomic-resolution probing of the perfect (1 × 1) surface provides the foundation for exploring the intrinsic sublattice features.

**Fe sublattice dichotomy**

We then focus on the small-bias tunneling spectra collected under liquid helium temperature to characterize the pairing gap. The normalized tunneling spectra along the two [100]-running cuts marked in Fig. 3(a) are plotted in Fig. 3(c) and the false-color plots in Fig. 3(d) (raw spectra in Fig. S4), i.e., along the α-Fe and $Se_-$ sites in the left panels and the β-Fe and $Se_+$ sites in the right panels. Figure 3(e) summarizes the typical spectra of the four atomic sites in one 2-Fe unit cell, taken at the same sites as Fig. 2(d). Consistent with previous works [1,20,22,29], the tunneling spectra present standard dual-gap structures, exhibiting sharp ± $V_i$ coherence peaks at ± 10 meV and broadened ± $V_o$ coherence peaks wiggled within an energy range of ± (15-17) meV. The outer gap variation is likely from local electronic modulation owing to the $SrTiO_3$ substrate [31]. In contrast, the inner coherence peaks are close to the $E_F$, resulting in a relatively small gap variation. Benefiting from the absence of electronic modulation, these tunneling spectra along each cut are almost homogeneous; thus, the lattice-dependent particle-hole asymmetric coherence peaks are discerned. Surprisingly, the spectra along these two cuts show reversed particle-hole asymmetry with opposite contrast in the dual gaps. Specifically, the spectra for α-Fe and $Se_-$ along the red dashed-line cut (left panels in Figs. 3(c) and 3(d)) consistently exhibit much higher inner-gap coherence peaks on the hole side $+V_i$ than on the electron side $-V_i$. In striking contrast, in the spectra for β-Fe and $Se_+$ along the blue dashed-line cut (right panels in Figs. 3(c) and 3(d)), the electron side of the inner coherence peak $-V_i$ becomes higher than the hole side $+V_i$. Moreover, such switched particle-hole asymmetries in $\pm V_i$ coherence peaks along the two cuts are associated with the reversal of particle-hole asymmetry in $\pm V_o$ coherence peaks. We define a ratio of the inner-gap coherence peak intensity $Z(r, V_i) = g(r, V_i)/g(r, -V_i)$ to quantify the particle-hole asymmetry and plot the ratios $Z(r, V_i)$ at $V_i$ = 10 meV for Fe sites in Fig. 3(b). The $Z(r, V_i)$ values are 1.1-1.5 for α-Fe and 0.6-0.9 for β-Fe.

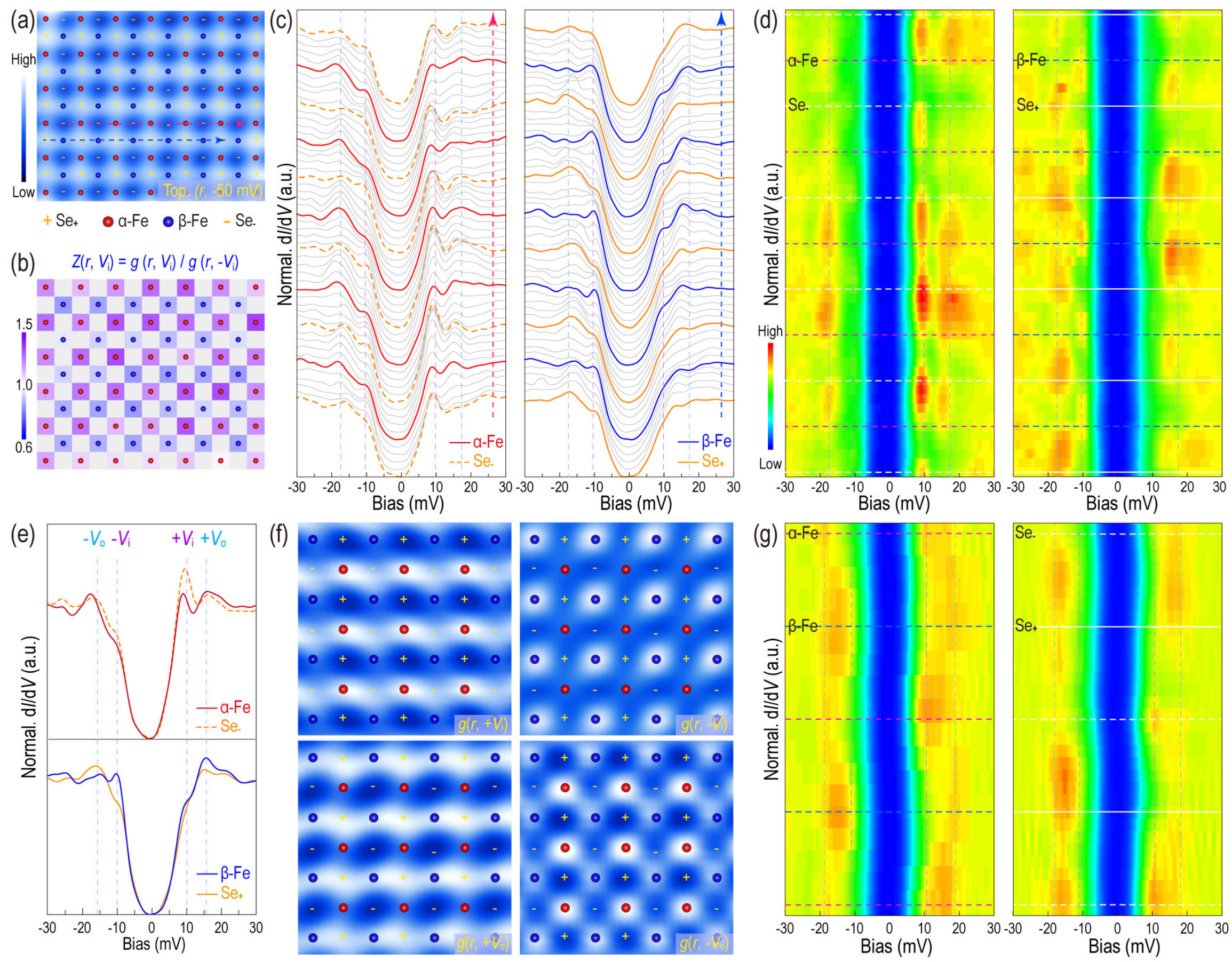

FIG 3. Fe-sublattice dichotomy in pairing state. (a) Atomic-resolution topography of the monolayer FeSe-(1×1) with the schematic of atomic sites overlaid. (b) Mosaic plots of $Z(r, V_i) = g(r, + V_i)/g(r, - V_i)$ for the Fe sites depicted in (a). $Z(r, V_i) > 1$ for α-Fe and $Z(r, V_i) < 1$ for β-Fe. (c) Normalized d$I$/d$V$ tunneling spectra taken along the [100]-running rows of α-Fe and $Se_-$ (left panel) marked by the red line and of β-Fe and $Se_+$ (right panel) marked by the blue line in (a), and (d) the false-color plots. The spectra are offset vertically for clarity. (e) Summarized typical tunneling spectra of the four atomic sites in a single 2-Fe unit cell. (f) Atomic-resolution $g(r, \pm V_i)$ and $g(r, \pm V_o)$ mapping images, showing sublattice intensity contrast: reversal between electron and hole sides of either gap and opposite between $V_i$ and $V_o$ in either side. (g) False-color plots of normalized d$I$/d$V$ tunneling spectra taken along the rows of Fe-Fe (left panel) and of $Se_+$-$Se_-$ (right panel). The vertical dashed lines in (c, d, e, g) are eye guides for the coherence peaks $\pm V_i$ and $\pm V_o$.

To further illustrate this sublattice-dependent pairing feature, we collect atomic-resolution $g(r, V)$ mapping images, specifically at the two pairs of coherence peaks $\pm V_i$ and $\pm V_o$. As exemplified in Fig. 3(f), the pairing gap exhibits periodic spatial modulation in the 2-Fe unit cell. In addition to the discernible contrasts between $Se_+$ and $Se_-$ sites, the two Fe sublattices exhibit universally more striking contrasts, particularly at $V = \pm V_i$. Regarding the tunneling matrix contrast between the $Se_+$ and $Se_-$ layers, it is more reasonable to focus on Fe-sites in the single layer. Compared with the β-Fe sites, the α-Fe sites consistently exhibit higher intensities at the hole side + $V_i$ and the electron side - $V_o$, but lower intensities at the electron side - $V_i$ and the hole side + $V_o$. That is, the Fe-sublattice dichotomy is featured by reversed particle-hole asymmetry with opposite contrast between the dual gaps. The false-color plot of the tunneling spectra along the Fe-Fe direction (left panel), in comparison with the plot along the Se-Se direction (right panel) in Fig. 3(g) (see details in Fig. S5), further reveals the distinct contrast between adjacent Fe sublattices. Moreover, the zero-conductance plateau centered at $E_F$ (in dark blue) shifts towards negative voltage for α-Fe, while oppositely towards positive voltage for β-Fe. Compared with the spectra variation for α-Fe and β-Fe, the spectra for $Se_-$ and $Se_+$ exhibit similar modulation, but with weakened contrasts in coherence peak heights and enhanced shifts in zero-conductance plateaus. The slightly stronger polar shift along the $Se_+$-$Se_-$ cut suggests a layer-oriented dipole effect, agreeing well with the interfacial charge transfer scenario. Therefore, the results shown in Fig. 3 completely present the

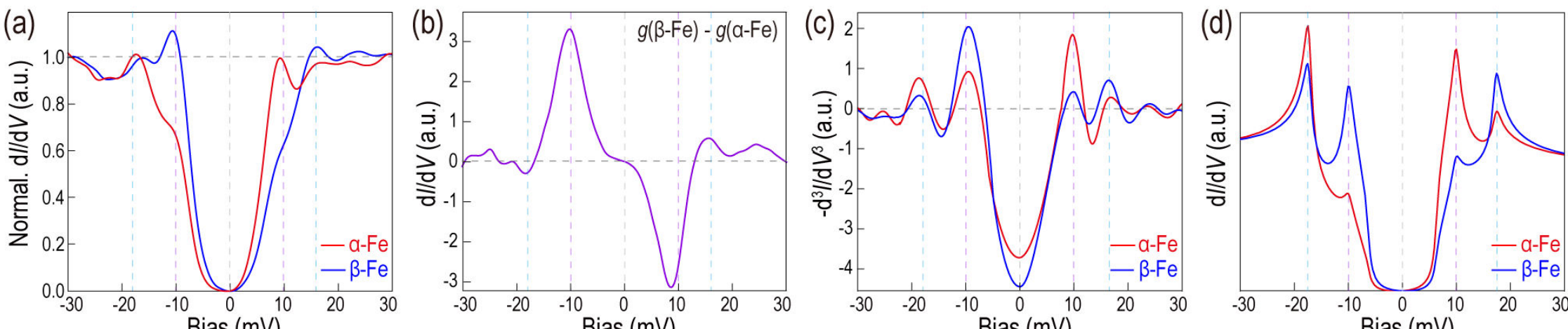

FIG 4. Sublattice dichotomy effect from mixed intraband pairing and interband pairing. (a) Two typical tunneling spectra at α-Fe and β-Fe showing sublattice dichotomy. (b) The difference in tunneling spectra between the two Fe sublattices. (c) Negative of the second derivative of the typical spectroscopy curves ($-d^3I/dV^3$) showing sublattice dichotomy. (d) Calculated tunneling DOS at two Fe-sites based on the $\boldsymbol{k} \cdot \boldsymbol{p}$ model with both normal pairing and interband pairing.

Fe-sublattice dichotomy illustrated in Fig. 1(c). Notably, the particle-hole asymmetry is independent of bias setpoint and spectra normalization (Section 3-2 of SM, Fig. S3). The sublattice dichotomy effect is reproducibly observed in the monolayer FeSe-(1×1) without any exception (Section 3-3 of SM, Figs. S4-S7), and even coexists with the (2×1) electronic modulation though with reduced contrast (Section 3-4 of SM, Figs. S8 and S9).

We briefly summarize our findings here. The distinct dual tunneling spectra corresponding to the two sublattices are exemplified by the typical tunneling spectra of α-Fe and β-Fe plotted in Fig. 4(a). For the inner-gap coherence peaks, α-Fe has a more pronounced hole peak at + $V_i$, whereas β-Fe has a more pronounced electron peak at - $V_i$. On the other hand, the outer-gap coherence peaks at ± $V_o$ show a reverse behavior to the inner-gap coherence peaks at ± $V_i$. More precisely, the + $V_i$ tunneling weight at α-Fe is almost equal to the - $V_i$ tunneling weight at β-Fe and vice versa. As revealed from the spectral difference between the two Fe sublattices, i.e., $g$(β-Fe)-$g$(α-Fe) displayed in Fig. 4(b), the sublattice contrasts appear only within the pairing gap ($|V_S| \leq 17\ mV$) and the tunneling spectra weights between the electron and hole sides are complementary, with an extreme contrast at ± $V_i$. This dual spectral behavior within the pairing gaps for the two Fe sublattices, in striking contrast to the minor monotonic variation observed in the normal state (Fig. 2(h) and Fig. S2), indicates an intrinsic property of the paired state. This is further manifested by the similar sublattice dichotomy features in the negative of the second derivative of the typical spectroscopy curves (- $\mathrm{d}^3I/\mathrm{d}V^3$) plotted in Fig. 4(c) (Section 3-5 of SM, Fig. S10), wherein the peaks serve as direct and sensitive indicators of the strength of superconducting coherence [32]. This unexpected intrinsic phenomenon points to a highly unusual pairing function inside the monolayer FeSe.

## IV. DISCUSSION

The existence of sublattice dichotomy implies broken space inversion symmetry. Indeed, as mentioned in the introduction, the space inversion symmetry is broken in the monolayer FeSe on $SrTiO_3$(001) [25-27]. However, the effect of symmetry breaking on electronic physics remains largely unclear and has been ignored in both theoretical modeling and experimental studies. The absence of inversion symmetry allows normal pairing and interband pairing to coexist. This odd-parity interband pairing is known to produce the two-gap feature [28]. Therefore, it is natural to inquire whether the combined action of interband pairing and normal pairing can produce the dichotomy effect [33].

To capture the aforementioned physics and monolayer FeSe electronic structure, we follow the $\boldsymbol{k} \cdot \boldsymbol{p}$ model around the $M$ point based on the symmetry of FeSe [34], and then, add both interband pairing and normal pairing into the $\boldsymbol{k} \cdot \boldsymbol{p}$ model. The local density of states (DOSs) at two sublattice sites are further calculated (see the model details in section 4-1 of SM). The tunneling DOSs for α-Fe and β-Fe are plotted in Fig. 4(d). Our theoretical simulation captures the sublattice dichotomy effect nicely. Besides the reversal of particle-hole asymmetry with opposite contrast between the dual gaps, the opposite shift in the tunneling spectrum around zero bias manifests the local particle-hole symmetry feature of interband pairing as well [33]. Notice that α-Fe and β-Fe are related by a $C_4$ symmetry centering at Se atoms. This $C_4$ symmetry is broken when interband pairing coexists with normal pairing, which is consistent with distinct DOSs on two sublattices (Section 4-3 of SM). Importantly, neither the normal pairing nor the interband pairing individually can produce the dichotomy effect (Section 4-2 of SM, Figs. S11(d, e)). It is also possible that the normal state already contains the symmetry-breaking effects. However, this normal state with normal pairing can always lead to sublattice differences, i.e., similar height differences of coherence peaks at the electron and hole sides between two Fe sublattices (Section 4-4 of SM, Fig. S11(g)), rather than particle-hole asymmetry. This is the scenario revealed very recently in exfoliated thin FeSe flakes [30]. In the case of the sublattice dichotomy we observed in the monolayer FeSe with broken inversion symmetry, interband pairing is an essential component.

## V. CONCLUSIONS

In summary, we achieve a systematic investigation of the sublattice degree of freedom of the monolayer FeSe with an exclusive (1 × 1) surface. Our atomically resolved tunneling spectra reveal dual-gap superconducting coherence peaks at $\pm V_{\mathrm{i}}$ and $\pm V_{\mathrm{o}}$ and dual spectral behaviors within the pairing gap energy for two Fe sublattices. More precisely, the coherence peak of α-Fe at $+V_{\mathrm{i}}$ is greater than that of β-Fe, while the coherence peak of β-Fe at $-V_{\mathrm{i}}$ is also greater than that of α-Fe by a comparable magnitude; whereas coherence peaks at $\pm V_{\mathrm{o}}$ exhibit reversed contrast. We have shown that the coexistence of the normal pairing between $\boldsymbol{k}$ and $-\boldsymbol{k}$ and the interband pairing between $\boldsymbol{k}$ and $-\boldsymbol{k} + \boldsymbol{Q}$ is the key mechanism for this sublattice dichotomy. This interband pairing is also an extension of the $\eta$-pairing classified in Ref. [35, 36].

The exceptionally high superconducting transition temperature in monolayer FeSe remains a central unresolved question in the field of iron-based superconductors. Deciphering the pairing structure in monolayer FeSe is key to addressing this mystery. Our research reveals a novel superconducting state in monolayer FeSe on $SrTiO_3$, placing significant constraints on possible pairing configurations, as well as pairing mechanisms. Specifically, our findings suggest that the $T_c$ enhancement arises from the activation of a new pairing channel, namely, interband pairing. Traditionally, interband pairing is not expected to dominate intraband pairing as the primary instability; however, our observation of sublattice dichotomy requires strong interband pairing, which cannot be simply understood as a conventional Fermi surface instability. The $T_c$ enhancement in monolayer FeSe has generated intriguing conjectures on pairing mechanisms, including interfacial electron-phonon enhancement [13,23,24] and incipient pairing mechanism [37-40]. However, these proposals do not specify such a large interband pairing component. Our results offer fresh insights into the pairing mechanisms in monolayer FeSe and underscore the complex and rich physics associated with sublattice degrees of freedom.

**ACKNOWLEDGMENTS**

The work is supported by the National Natural Science Foundation of China (Grant Nos. 92477204, 52388201, 1888101, 12174428, 12404157 and 11920101005), the National Key Research and Development Program of China (Grant Nos. 2022YFA1403100 and 2022YFA1403900), the Strategic Priority Research Program of the Chinese Academy of Sciences (Grant Nos. XDB28000000 and XDB33000000), and the New Cornerstone Investigator Program, the Chinese Academy of Sciences Project for Young Scientists in Basic Research (2022YSBR-048).